\documentclass[twocolumn,showpacs,preprintnumbers,amsmath,amssymb]{revtex4}
\usepackage{natbib}
\usepackage{lgrind}
\usepackage{latexcad}
\usepackage{float}
\usepackage{amssymb}
\usepackage{amsmath,amsfonts}
\usepackage{graphicx}
\usepackage{rotating}
\usepackage{dcolumn}
\usepackage{mathrsfs}
\usepackage{bm}
\begin{document}

\baselineskip=6.3mm

%\preprint{NLP/0401-SJTU}

\title{Time-Periodic Solutions of the Einstein's Field Equations II}
\author{De-Xing Kong$^{1}$, Kefeng Liu$^{2,3}$ and Ming Shen$^2$}
%and Qing-You Sun$^2$}
\affiliation{\\
$^{1}$\!Department of Mathematics,
Zhejiang University, Hangzhou 310027, China\\
$^{2}$\!Center of Mathematical Sciences,
Zhejiang University, Hangzhou 310027, China\\
$^3$\!Department of Mathematics, University of California at Los
Angeles, CA 90095, USA}
\date{\today}% It is always \today, today,

\begin{abstract}
In this paper, we construct several kinds of new time-periodic
solutions of the vacuum Einstein's field equations whose Riemann
curvature tensors vanish, keep finite or take the infinity at some
points in these space-times, respectively. The singularities of
these new time-periodic solutions are investigated and some new
physical phenomena are found. The applications of these solutions in
modern cosmology and general relativity can be expected.
\end{abstract}


%Use showkeys class option if keyword

\maketitle

{\em 1. Introduction.} The Einstein's field equations are the
fundamental equations in general relativity and play an essential
role in cosmology. This paper concerns the time-periodic solutions
of the following vacuum Einstein's field equations
\begin{equation}
G_{\mu\nu}\stackrel{\triangle}{=}R_{\mu\nu}-\frac{1}{2}g_{\mu\nu}R=0,
\end{equation}
or equivalently,
\begin{equation}
R_{\mu\nu}=0,
\end{equation}
where $g_{\mu\nu}\;(\mu,\nu=0,1,2,3)$ is the unknown Lorentzian
metric, $R_{\mu\nu}$ is the Ricci curvature tensor, $R$ is the
scalar curvature and $G_{\mu\nu}$ is the Einstein tensor.

It is well known that the exact solutions of the Einstein's field
equations play a crucial role in general relativity and cosmology.
Typical examples are the Schwarzschild solution and Kerr solution.
Although many interesting and important solutions have been obtained
(see, e.g., \cite{b} and \cite{skmhh}), there are still many
fundamental open problems. One such problem is {\it if there exists
a ``time-periodic" solution, which contains physical singularities
such as black hole, to the Einstein's field equations}. This paper
continues the discussion of this problem.

The first time-periodic solution of the vacuum Einstein's field
equations was constructed by the first two authors in \cite{kl}. The
solution presented in \cite{kl} is time-periodic, and describes a
regular space-time, which has vanishing Riemann curvature tensor but
is inhomogenous, anisotropic and not asymptotically flat. In
particular, this space-time does not contain any essential
singularity, but contains some non-essential singularities which
correspond to steady event horizons, time-periodic event horizon and
has some interesting new physical phenomena.

In this paper, we focus on finding the time-periodic solutions,
which contain physical singularities such as black hole to the
vacuum Einstein's field equations (1). We shall construct three
kinds of new time-periodic solutions of the vacuum Einstein's field
equations (1) whose Riemann curvature tensors vanish, keep finite or
go to the infinity at some points in these space-times respectively.
The singularities of these new time-periodic solutions are
investigated and new physical phenomena are found. Moreover, the
applications of these solutions in modern cosmology and general
relativity may be expected. In the forthcoming paper \cite{kl2}, we
shall construct a time-periodic solution of the Einstein's field
equations with black hole, which describes the time-periodic
cosmology with many new and interesting physical phenomena.

{\em 2. Procedure of finding new solutions.}

We consider the metric of the following form
\begin{equation}\left(g_{\mu\nu}\right)=\left( \begin{array}{cccc}
u & v & p & 0 \\
v & 0 & 0 & 0 \\
p & 0 & f & 0 \\
0 & 0 & 0 & h \\
\end{array}\right),\end{equation}
where $u,v,p,f$ and $h$ are smooth functions of the coordinates
$(t,x,y,z)$. It is easy to verify that the determinant of
$(g_{\mu\nu})$ is given by
\begin{equation}g\stackrel{\triangle}{=}\det(g_{\mu\nu})=
-{v}^{2}fh. \end{equation}

Throughout this paper, we assume that
$$g<0.\eqno{(H)}$$
Without loss of generality, we may suppose that $f$ and $g$ keep the
same sign, for example,
\begin{equation}
f < 0\;\; (resp.\; f>0)\quad{\rm{and}}\quad h < 0 \;\; (resp.\;
g>0).\end{equation}

In what follows, we solve the Einstein's field equations (2) under
the framework of the Lorentzian metric of the form (3).

By a direct calculation, we have the Ricci tensor
\begin{equation}\begin{array}{lll}
R_{11} & = & {\displaystyle
-\frac{1}{2}\left\{\frac{v_x}{v}\left(\frac{f_x}{f}+\frac{h_x}{h}\right)+\right.}\vspace{2mm}\\
& &  {\displaystyle\left.
\frac{1}{2}\left[\left(\frac{f_x}{f}\right)^2+\left(\frac{h_x}{h}\right)^2\right]-\left(
\frac{f_{xx}}{f}+\frac{h_{xx}}{h}\right)\right\}.}\end{array}
\end{equation}
It follows from (2) that
\begin{equation}
\frac{v_x}{v}\left(\frac{f_x}{f}+\frac{h_x}{h}\right)+
\frac{1}{2}\left[\left(\frac{f_x}{f}\right)^2+\left(\frac{h_x}{h}\right)^2\right]-\left(
\frac{f_{xx}}{f}+\frac{h_{xx}}{h}\right)=0.
\end{equation}
This is an ordinary differential equation of first order on the
unknown function $v$. Solving (7) gives
\begin{equation}
v = V(t,y,z)\exp\left\{\int \Theta(t,x,y,z)dx\right\},
\end{equation}
where
$$\Theta =
\left[\displaystyle\frac{f_{xx}}{f}+\displaystyle\frac{h_{xx}}{h}-
\displaystyle\frac{1}{2}\left(\displaystyle\frac{f_x}{f}\right)^2-
\displaystyle\frac{1}{2}\left(\displaystyle\frac{h_x}{h}\right)^2\right]
 \displaystyle\frac{fh}{(fh)_x},$$
and $V=V(t,y,z)$ is an integral function depending on $t$, $y$ and
$z$. Here we assume that
\begin{equation}(fh)_x\neq 0.\end{equation}
In particular, taking the ansatz
\begin{equation}
f=-K(t,x)^2,\quad h=N(t,y,z)K(t,x)^2
\end{equation}
and substituting it into (8) yields
\begin{equation}
v=VK_x.
\end{equation}
By the assumptions (H) and (9), we have
\begin{equation}
V \neq 0, \quad K\neq 0, \quad K_x\neq 0.
\end{equation}

Noting (10) and (11), by a direct calculation we obtain
\begin{equation}
R_{13}=-\frac{V_zK_x}{KV}.
\end{equation}
It follows from (2) that
$$R_{13}=0.$$
Combining (12) and (13) gives
\begin{equation}
V_z=0.
\end{equation}
This implies that the function $V$ depends only on $t,\;y$ but is
independent of $x$ and $z$. Noting (10)-(11) and using (14), we
calculate
\begin{equation}
R_{12}=-\frac{1}{2V}\left(\displaystyle\frac{p_{xx}}{K_x}-\displaystyle\frac{K_{xx}p_x}
{K_x^2}-\displaystyle\frac{2pK_x}{K^2}+\displaystyle\frac{2K_xV_y}{K}\right).
\end{equation}
Solving $p$ from the equation $R_{12}=0$ yields
\begin{equation}
p=AK^2+V_yK+\displaystyle\frac{B}{K},
\end{equation}
where $A$ and $B$ are integral functions depending on $t$, $y$ and
$z$. Noting (10)-(11) and using (14) and (16), we observe that the
equation $R_{23}=0$ is equivalent to
\begin{equation}
 B_z-2K^3A_z=0.
\end{equation}
Since $K$ is a function depending only on $t,\; x$, and $A,\; B$ are
functions depending on $t,\;y$ and $z$, we can obtain that
\begin{equation}
B=2K^3A+C(t,x,y),
\end{equation}
where $C$ is an integral function depending on $t,\;x$ and $y$. For
simplicity, we take
\begin{equation}
A=B=C=0.
\end{equation}
Thus, (16) simplifies to
\begin{equation}
p=V_yK.
\end{equation}

From now on, we assume that the function $N$ only depends on $y$,
that is to say,
\begin{equation}
 N=N(y).
\end{equation}
Substituting (10)-(11), (14) and (20)-(21) into the equation
$R_{02}=0$ yields
\begin{equation}
u_xV_y+V(u_{yx}-4V_yK_{xt})=0.
\end{equation}
Solving $u$ from the equation (22) leads to
\begin{equation}
u=2K_tV.
\end{equation}
Noting (10)-(11), (14), (20)-(21) and (23), by a direct calculation
we obtain
\begin{equation}
R_{03}=0,
\end{equation}
\begin{equation}\left\{\begin{array}{lll}
R_{22} & = & (4N^2V^2)^{-1}\left[2NV^2N_{yy}-4N^2VV_{yy}\right.\vspace{2mm}\\
& &\left. \qquad +4N^2V_y^2-2NVN_yV_y-V^2N_y^2 \right],\vspace{2mm}\\
R_{33} & = & -(4NV^2)^{-1}\left[2NV^2N_{yy}-4N^2VV_{yy}\right.\vspace{2mm}\\
& &\left. \qquad +4N^2V_y^2-2NVN_yV_y-V^2N_y^2 \right]
\end{array}\right.
\end{equation}
and
\begin{equation}\begin{array}{lll}
R_{00} & = & (2KNV^2)^{-1}\left[4NV_tV_y^2+2NV^2V_{tyy}-2NVV_tV_{yy}\right.\vspace{2mm}\\
& &\left. \qquad
-4NVV_yV_{ty}-VN_yV_tV_y+V^2N_yV_{ty}\right].\end{array}
\end{equation}
Therefore, under the assumptions mentioned above, the Einstein's
field equations (2) are reduced to
\begin{equation}
-\frac{N_{yy}}{N}+\frac{1}{2}\left(\frac{N_y}{N}\right)^2
+2\frac{V_{yy}}{V}+\frac{N_yV_y}{NV}-2\left(\frac{V_y}{V}\right)^2=0
\end{equation}
and
\begin{equation}\begin{array}{rrr}
& {\displaystyle 4V_y^2V_t+2V^2V_{yyt}-2VV_{yy}V_t-4VV_yV_{yt}} & \vspace{3mm}\\
 &  {\displaystyle -
\frac{VV_yV_tN_y}{N}+\frac{V^2V_{yt}N_y}{N}} &  =0.\end{array}
\end{equation}
On the other hand, (27) can be rewritten as
\begin{equation}
2\left(\frac{V_y}{V}\right)_y+\frac{V_yN_y}{VN}-\left(\frac{N_y}{N}\right)_y-\frac{1}{2}\left(\frac{N_y}{N}\right)^2=0
\end{equation}
and (28) is equivalent to
\begin{equation}
2\left(\frac{V_y}{V}\right)_{yt}+\left(\frac{V_y}{V}\right)_t\frac{N_y}{N}=0.
\end{equation}
Noting (21) and differentiating (29) with respect to $t$ gives (30)
directly. This shows that (29) implies (30). Hence in the present
situation, the Einstein's field equations (2) are essentially (29).
Solving $V$ from the equation (29) yields
\begin{equation}
V=w(t)|N(y)|^{1/2}\exp\left\{q(t)\int |N(y)|^{-1/2}dy\right\},
\end{equation}
where $w=w(t)$ and $q=q(t)$ are two integral functions only
depending on $t$. Thus, we can obtain the following solution of the
vacuum Einstein's field equations in the coordinates $(t,x,y,z)$
\begin{equation} \label{eq:4.6}
ds^2=(dt,dx,dy,dz)(g_{\mu\nu})(dt,dx,dy,dz)^T,
\end{equation}
where
\begin{equation}\label{eq:4.7}(g_{\mu\nu})=\left( \begin{array}{cccc}
2K_tV & K_xV& KV_y & 0\\K_xV  & 0 & 0 & 0 \\
 KV_y& 0 & -K^2 & 0\\
0 & 0 & 0 & NK^2\\
\end{array}\right),\end{equation}
in which $N=N(y)$ is an arbitrary function of $y$, $K=K(t,x)$ is an
arbitrary function of $t,\;x$, and $V$ is given by (31).

By calculations, the Riemann curvature tensor reads
\begin{equation}
R_{\alpha\beta\mu\nu}=0,\quad\forall \; \alpha\beta\mu\nu\neq 0202
\;\; {\rm or}\;\; 0303,
\end{equation}
while
\begin{equation}
R_{0202}=Kwqq'|N|^{-1/2}\exp\left\{q\int|N|^{-1/2}dy\right\}
\end{equation}
and
\begin{equation}
R_{0303}=Kwqq'|N|^{1/2}\exp\left\{q\int|N|^{-1/2}dy\right\}.
\end{equation}

{\em 3. Time-periodic solutions.} This section is devoted to
constructing some new time-periodic solutions of the vacuum
Einstein's field equations.

{\em 3.1 Regular time-periodic space-times with vanishing Riemann
curvature tensor.} Take $q=constant$ and let $V=\rho(t)\kappa(y)$,
where $\kappa$ is defined by
\begin{equation}
\kappa(y)=c_1\sqrt{|N|}\exp\left\{c_2\int|N|^{-1/2}dy\right\},
\end{equation}
in which $c_1$ and $c_2$ are two integrable constants. In this case,
the solution to the vacuum Einstein's filed equations in the
coordinates $(t,x,y,z)$ reads
\begin{equation} \label{eq:4.6}
ds^2=(dt,dx,dy,dz)(g_{\mu\nu})(dt,dx,dy,dz)^T,
\end{equation}
where
\begin{equation} \label{eq:4.7}
(g_{\mu\nu})= \left(
\begin{array}{cccc}
2\rho\kappa\partial_tK &\rho\kappa\partial_xK &\rho K\partial_y\kappa & 0 \\
\rho\kappa\partial_xK &0  & 0 & 0  \\
\rho K\partial_y \kappa & 0 & -K^2 & 0 \\
0 & 0 & 0 & NK^2 \\
\end{array}\right).
\end{equation}

\noindent{\bf Theorem 1} {\em The vacuum Einstein's filed equations
(2) have a solution described by (38) and (39), and the Riemann
curvature tensor of this solution vanishes.}$\quad\blacksquare$

As an example, let
\begin{equation}\left\{\begin{array}{l}
w(t)=\cos t,\vspace{2mm}\\
q(t)=0,\vspace{2mm}\\
 K(t,x)=e^x\sin t,\vspace{2mm}\\
  N(y)=-(2+\sin y)^2.\end{array}\right.\end{equation}
In the present situation, we obtain the following solution of the
vacuum Einstein's filed equations (2)
\begin{equation}
(\eta_{\mu\nu})=\left( \begin{array}{cccc}
\eta_{00} & \eta_{01} & \eta_{02} & 0\\ \eta_{01} & 0 & 0 & 0 \\
\eta_{02} & 0 & \eta_{22} & 0\\
0 & 0 & 0 & \eta_{33}\\
\end{array}\right),\end{equation}
where
\begin{equation}
\left\{\begin{array}{l}
\eta_{00}=2e^x(2+\sin y)\cos^2t,\vspace{2mm}\\
\eta_{01}={\displaystyle \frac12 e^x (2+\sin y)\sin (2t)},\vspace{2mm}\\
\eta_{02}={\displaystyle \frac 12 e^x\cos y\sin(2t),}\vspace{2mm}\\
\eta_{22}=-[e^x\sin t]^2,\vspace{2mm}\\
\eta_{33}=-[e^x (2+\sin y)\sin t]^2.
\end{array}\right.
\end{equation}
By (4),
\begin{equation}\eta\stackrel{\triangle}{=}\det(\eta_{\mu\nu})=-\frac14 e^{6x}
(2+\sin y)^4\sin^4t\sin^2(2t). \end{equation}

\noindent{\bf Property 1} The solution (41) of the vacuum Einstein's
filed equations (2) is time-periodic.$\quad\blacksquare$

\noindent{\bf Proof.} In fact, the first equality in (42) implies
that $$ \eta_{00} >0 \quad {\rm for}\;\; t\neq k\pi+\pi/2\;\; (k\in
\mathbb{N})\;\;{\rm and}\;\; x\neq-\infty.$$ On the other hand, by
direct calculations,
$$
\left|\begin{array}{cccc}
\eta_{00} & \eta_{01} \\
\eta_{01}& 0
\end{array}\right|=-\frac14 e^{2x}(2+\sin y)^2\sin^2(2t)<0,$$
$$\left|\begin{array}{cccc}
\eta_{00} & \eta_{01}&\eta_{02}\\
\eta_{01} & 0& 0 \\
\eta_{02}& 0 &\eta_{22}
\end{array}\right|=-\eta_{01}^2\eta_{22}
>0$$
and
$$\left|\begin{array}{cccc}
\eta_{00}& \eta_{01} &\eta_{02} & 0  \\
\eta_{01} & 0 & 0 & 0  \\
\eta_{02} & 0& \eta_{22} & 0 \\
0 & 0 & 0 & \eta_{33}
\end{array}\right|=-\eta_{01}^2\eta_{22}\eta_{33}<0$$
for $t\neq k\pi,\;k\pi+\pi/2\;\;(k\in \mathbb{N})$ and
$x\neq-\infty$.

In Property 3 below, we will show that $t=
k\pi,\;k\pi+\pi/2\;\;(k\in \mathbb{N})$ are the singularities of the
space-time described by (41), but they are not essential (or say,
physical) singularities, these non-essential singularities
correspond to the event horizons of the space-time described by (41)
with (42); while, when $x=-\infty$, the space-time (41) degenerates
to a point.

The above discussion implies that the variable $t$ is a time
coordinate. Therefore, it follows from (42) that the Lorentzian
metric
\begin{equation}
ds^2=(dt,dx,dy,dz)(\eta_{\mu\nu})(dt,dx,dy,dz)^T
\end{equation}
is indeed a time-periodic solution of the vacuum Einstein's field
equations (2), where $(\eta_{\mu\nu})$ is given by (41). This proves
Property 1. $\quad\quad\quad\square$

Noting (34)-(36) and the second equality in (40) gives

\noindent{\bf Property 2} The Lorentzian metric (44) (in which
$(\eta_{\mu\nu})$ is given by (41) and (42)) describes a regular
space-time, this space-time is Riemannian flat, that is to say, its
Riemann curvature tensor vanishes. $\quad\blacksquare$

\noindent{\bf Remark 1} {\em The first time-periodic solution to the
Einstein's field equations was constructed by Kong and Liu
\cite{kl}. The time-periodic solution presented in \cite{kl} also
has the vanishing Riemann curvature tensor.}

It follows from (43) that the hypersurfaces $t= k\pi$,
$k\pi+\pi/2\;\;(k\in \mathbb{N})$ and $x=\pm\infty$ are
singularities of the space-time (44) (in which $(\eta_{\mu\nu})$ is
given by (41) and (42)), however, by Property 2, these singularities
are not physical (or say, not essential). According to the
definition of event horizon (see e.g., Wald \cite{wald}), it is easy
to show that the hypersurfaces $t= k\pi,\;k\pi+\pi/2\;\;(k\in
\mathbb{N})$ and $x=+\infty$ are the event horizons of the
space-time (44) (in which $(\eta_{\mu\nu})$ is given by (41) and
(42)). Therefore, we have

\noindent{\bf Property 3}  The Lorentzian metric (44) (in which
$(\eta_{\mu\nu})$ is given by (41) and (42)) does not contain any
essential singularity. These non-essential singularities consist of
the hypersurfaces $t= k\pi,\;k\pi+\pi/2\;\;(k\in \mathbb{N})$ and
$x=\pm\infty$. The singularities $t= k\pi,\;k\pi+\pi/2\;\;(k\in
\mathbb{N})$ and $x=+\infty$ correspond to the event horizons,
while, when $x=-\infty$, the space-time (44) degenerates to a point.
$\quad\blacksquare$

We now investigate the physical behavior of the space-time (44).

Fixing $y$ and $z$, we get the induced metric
\begin{equation}
ds^2=\eta_{00}dt^2+2\eta_{01}dtdx.
\end{equation}
Consider the null curves in the $(t,x)$-plan, which are defined by
\begin{equation}
\eta_{00}dt^2+2\eta_{01}dtdx=0.
\end{equation}
Noting (42) gives
\begin{equation}
dt=0 \quad {\rm and} \quad \frac{dt}{dx}=-\tan t.
\end{equation}
Thus, the null curves and light-cones are shown in Figure 1.
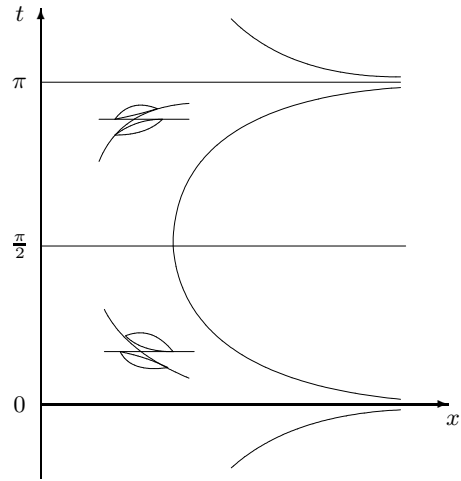
\begin{figure}[H]
    \begin{center}
\begin{picture}(190,186)
\thinlines \drawvector{26.0}{4.0}{178.0}{0}{1}
\drawpath{26.0}{92.0}{164.0}{92.0}
\drawpath{26.0}{154.0}{162.0}{154.0}
\path(76.0,92.0)(76.0,92.0)(76.0,93.05)(76.01,94.12)(76.05,95.19)(76.12,96.23)(76.19,97.26)(76.3,98.3)(76.41,99.3)(76.55,100.33)
\path(76.55,100.33)(76.69,101.33)(76.86,102.3)(77.01,103.29)(77.22,104.26)(77.44,105.22)(77.66,106.16)(77.91,107.11)(78.19,108.05)(78.47,108.97)
\path(78.47,108.97)(78.76,109.87)(79.08,110.76)(79.44,111.68)(79.76,112.55)(80.16,113.41)(80.51,114.3)(80.94,115.12)(81.36,116.0)(81.8,116.83)
\path(81.8,116.83)(82.25,117.66)(82.73,118.47)(83.22,119.26)(83.73,120.08)(84.25,120.86)(84.8,121.62)(85.36,122.41)(85.94,123.16)(86.51,123.91)
\path(86.51,123.91)(87.12,124.62)(87.76,125.37)(88.41,126.08)(89.08,126.8)(89.75,127.5)(90.44,128.21)(91.16,128.88)(91.87,129.55)(92.62,130.22)
\path(92.62,130.22)(93.41,130.86)(94.19,131.52)(94.98,132.13)(95.8,132.77)(96.62,133.38)(97.48,133.99)(98.36,134.58)(99.25,135.16)(100.12,135.75)
\path(100.12,135.75)(101.05,136.32)(102.0,136.86)(102.94,137.41)(103.94,137.96)(104.91,138.49)(105.91,139.0)(106.94,139.5)(108.0,140.0)(109.05,140.5)
\path(109.05,140.5)(110.12,140.97)(111.22,141.44)(112.33,141.91)(113.44,142.36)(114.58,142.8)(115.75,143.24)(116.94,143.66)(118.12,144.07)(119.33,144.47)
\path(119.33,144.47)(120.58,144.86)(121.8,145.25)(123.08,145.63)(124.36,145.99)(125.66,146.35)(126.97,146.69)(128.3,147.02)(129.66,147.36)(131.01,147.66)
\path(131.01,147.66)(132.41,147.97)(133.8,148.27)(135.23,148.57)(136.66,148.85)(138.13,149.11)(139.6,149.36)(141.08,149.61)(142.58,149.86)(144.11,150.08)
\path(144.11,150.08)(145.63,150.3)(147.21,150.52)(148.77,150.72)(150.38,150.91)(151.97,151.1)(153.61,151.27)(155.25,151.44)(156.91,151.58)(158.58,151.74)
\path(158.58,151.74)(160.27,151.86)(161.99,152.0)(162.0,152.0)
\path(76.0,92.0)(76.0,92.0)(76.08,91.0)(76.19,90.01)(76.3,89.01)(76.44,88.05)(76.58,87.08)(76.75,86.15)(76.94,85.19)(77.12,84.26)
\path(77.12,84.26)(77.33,83.33)(77.55,82.41)(77.8,81.5)(78.08,80.58)(78.33,79.69)(78.62,78.8)(78.94,77.94)(79.26,77.05)(79.61,76.19)
\path(79.61,76.19)(79.94,75.36)(80.33,74.51)(80.72,73.68)(81.11,72.83)(81.51,72.01)(81.94,71.22)(82.41,70.41)(82.86,69.62)(83.33,68.83)
\path(83.33,68.83)(83.83,68.05)(84.33,67.29)(84.86,66.51)(85.41,65.76)(85.97,65.01)(86.51,64.3)(87.12,63.56)(87.72,62.84)(88.33,62.13)
\path(88.33,62.13)(88.97,61.43)(89.62,60.72)(90.3,60.04)(90.97,59.36)(91.66,58.72)(92.37,58.04)(93.11,57.4)(93.86,56.75)(94.61,56.11)
\path(94.61,56.11)(95.37,55.5)(96.16,54.86)(96.98,54.25)(97.8,53.65)(98.62,53.08)(99.48,52.5)(100.36,51.9)(101.25,51.34)(102.12,50.79)
\path(102.12,50.79)(103.05,50.22)(103.98,49.68)(104.94,49.15)(105.87,48.63)(106.86,48.11)(107.86,47.61)(108.86,47.11)(109.87,46.61)(110.94,46.13)
\path(110.94,46.13)(111.98,45.65)(113.05,45.2)(114.12,44.74)(115.25,44.29)(116.36,43.84)(117.5,43.4)(118.62,42.99)(119.8,42.58)(120.98,42.15)
\path(120.98,42.15)(122.19,41.77)(123.37,41.38)(124.62,40.99)(125.86,40.61)(127.12,40.25)(128.38,39.9)(129.69,39.54)(130.98,39.2)(132.3,38.88)
\path(132.3,38.88)(133.63,38.54)(135.0,38.24)(136.36,37.93)(137.75,37.63)(139.13,37.34)(140.55,37.06)(141.99,36.77)(143.44,36.52)(144.88,36.25)
\path(144.88,36.25)(146.36,36.02)(147.86,35.77)(149.36,35.54)(150.88,35.31)(152.44,35.11)(153.99,34.9)(155.55,34.7)(157.13,34.5)(158.75,34.33)
\path(158.75,34.33)(160.36,34.15)(161.99,34.0)(162.0,34.0)
\path(98.0,178.0)(98.0,178.0)(98.44,177.55)(98.87,177.11)(99.33,176.69)(99.79,176.27)(100.25,175.85)(100.69,175.41)(101.16,175.02)(101.62,174.61)
\path(101.62,174.61)(102.12,174.21)(102.58,173.82)(103.08,173.41)(103.55,173.02)(104.05,172.63)(104.55,172.27)(105.05,171.88)(105.55,171.52)(106.05,171.13)
\path(106.05,171.13)(106.55,170.77)(107.08,170.41)(107.58,170.08)(108.12,169.72)(108.62,169.38)(109.16,169.02)(109.69,168.69)(110.25,168.36)(110.79,168.02)
\path(110.79,168.02)(111.33,167.72)(111.87,167.38)(112.44,167.08)(113.0,166.77)(113.55,166.47)(114.12,166.16)(114.69,165.86)(115.26,165.58)(115.83,165.27)
\path(115.83,165.27)(116.43,165.0)(117.01,164.72)(117.58,164.44)(118.19,164.16)(118.79,163.91)(119.4,163.63)(120.0,163.38)(120.61,163.13)(121.23,162.88)
\path(121.23,162.88)(121.83,162.63)(122.47,162.41)(123.08,162.16)(123.72,161.94)(124.36,161.72)(124.98,161.5)(125.62,161.27)(126.26,161.05)(126.93,160.85)
\path(126.93,160.85)(127.58,160.63)(128.23,160.44)(128.91,160.25)(129.55,160.05)(130.23,159.88)(130.91,159.69)(131.58,159.52)(132.26,159.33)(132.94,159.16)
\path(132.94,159.16)(133.64,159.0)(134.33,158.85)(135.04,158.69)(135.75,158.52)(136.44,158.38)(137.16,158.25)(137.88,158.11)(138.58,157.97)(139.32,157.85)
\path(139.32,157.85)(140.02,157.72)(140.75,157.6)(141.5,157.47)(142.22,157.36)(142.97,157.25)(143.72,157.16)(144.47,157.05)(145.22,156.97)(145.97,156.88)
\path(145.97,156.88)(146.75,156.77)(147.5,156.71)(148.27,156.63)(149.07,156.55)(149.83,156.49)(150.63,156.41)(151.41,156.36)(152.19,156.3)(153.0,156.25)
\path(153.0,156.25)(153.77,156.22)(154.6,156.16)(155.38,156.13)(156.21,156.1)(157.02,156.07)(157.83,156.05)(158.66,156.02)(159.49,156.0)(160.32,156.0)
\path(160.32,156.0)(161.16,156.0)(161.99,156.0)(162.0,156.0)
\path(162.0,30.0)(162.0,30.0)(161.16,29.95)(160.32,29.88)(159.49,29.86)(158.66,29.79)(157.83,29.75)(157.02,29.69)(156.21,29.62)(155.38,29.54)
\path(155.38,29.54)(154.6,29.46)(153.8,29.42)(153.0,29.34)(152.19,29.26)(151.41,29.17)(150.63,29.05)(149.83,28.96)(149.07,28.87)(148.27,28.79)
\path(148.27,28.79)(147.5,28.69)(146.75,28.59)(146.0,28.45)(145.22,28.36)(144.47,28.21)(143.72,28.12)(142.97,28.0)(142.25,27.87)(141.5,27.71)
\path(141.5,27.71)(140.75,27.6)(140.02,27.45)(139.32,27.29)(138.58,27.18)(137.88,27.03)(137.16,26.87)(136.44,26.7)(135.75,26.54)(135.05,26.37)
\path(135.05,26.37)(134.33,26.2)(133.64,26.04)(132.94,25.87)(132.26,25.7)(131.58,25.52)(130.91,25.3)(130.23,25.12)(129.55,24.95)(128.91,24.75)
\path(128.91,24.75)(128.25,24.54)(127.58,24.35)(126.93,24.12)(126.26,23.93)(125.62,23.7)(125.0,23.5)(124.36,23.27)(123.72,23.04)(123.08,22.79)
\path(123.08,22.79)(122.47,22.59)(121.83,22.35)(121.23,22.11)(120.61,21.87)(120.0,21.62)(119.4,21.37)(118.8,21.12)(118.19,20.86)(117.58,20.6)
\path(117.58,20.6)(117.01,20.3)(116.43,20.04)(115.83,19.79)(115.26,19.51)(114.69,19.2)(114.12,18.95)(113.55,18.67)(113.0,18.37)(112.44,18.05)
\path(112.44,18.05)(111.87,17.78)(111.33,17.45)(110.79,17.18)(110.25,16.87)(109.69,16.54)(109.16,16.21)(108.62,15.92)(108.12,15.6)(107.58,15.27)
\path(107.58,15.27)(107.08,14.93)(106.55,14.6)(106.05,14.27)(105.55,13.93)(105.05,13.59)(104.55,13.22)(104.05,12.89)(103.55,12.52)(103.08,12.18)
\path(103.08,12.18)(102.58,11.81)(102.12,11.43)(101.62,11.06)(101.16,10.71)(100.69,10.31)(100.25,9.93)(99.79,9.56)(99.33,9.18)(98.87,8.77)
\path(98.87,8.77)(98.44,8.39)(98.0,8.0)(98.0,8.0)
\drawcenteredtext{18.0}{180.0}{$t$}
\drawcenteredtext{18.0}{154.0}{$\pi$}
\drawcenteredtext{18.0}{92.0}{$\frac{\pi}{2}$}
\drawcenteredtext{18.0}{32.0}{$0$}
\drawvector{26.0}{32.0}{154.0}{1}{0}
\drawcenteredtext{182.0}{26.0}{$x$}
\path(48.0,124.0)(48.0,124.0)(48.15,124.37)(48.31,124.79)(48.47,125.18)(48.65,125.55)(48.84,125.94)(49.02,126.33)(49.2,126.69)(49.38,127.08)
\path(49.38,127.08)(49.58,127.44)(49.77,127.8)(49.97,128.16)(50.15,128.52)(50.36,128.88)(50.59,129.24)(50.79,129.58)(51.02,129.91)(51.22,130.27)
\path(51.22,130.27)(51.45,130.61)(51.68,130.94)(51.9,131.27)(52.15,131.6)(52.38,131.91)(52.61,132.24)(52.86,132.55)(53.11,132.86)(53.36,133.16)
\path(53.36,133.16)(53.61,133.47)(53.88,133.77)(54.15,134.08)(54.4,134.38)(54.68,134.66)(54.95,134.94)(55.24,135.22)(55.52,135.5)(55.79,135.77)
\path(55.79,135.77)(56.09,136.05)(56.38,136.33)(56.65,136.6)(56.97,136.86)(57.27,137.11)(57.58,137.36)(57.88,137.61)(58.2,137.86)(58.52,138.11)
\path(58.52,138.11)(58.84,138.35)(59.15,138.58)(59.49,138.82)(59.81,139.05)(60.15,139.27)(60.49,139.5)(60.84,139.71)(61.18,139.91)(61.52,140.13)
\path(61.52,140.13)(61.88,140.35)(62.24,140.55)(62.59,140.75)(62.95,140.94)(63.33,141.13)(63.7,141.33)(64.05,141.5)(64.44,141.69)(64.83,141.88)
\path(64.83,141.88)(65.22,142.05)(65.61,142.22)(66.0,142.38)(66.4,142.55)(66.79,142.71)(67.19,142.86)(67.58,143.02)(68.0,143.16)(68.43,143.32)
\path(68.43,143.32)(68.83,143.46)(69.27,143.6)(69.68,143.74)(70.11,143.86)(70.54,144.0)(70.99,144.11)(71.43,144.24)(71.86,144.36)(72.3,144.47)
\path(72.3,144.47)(72.75,144.58)(73.22,144.69)(73.68,144.77)(74.12,144.88)(74.58,144.99)(75.05,145.08)(75.54,145.16)(76.0,145.25)(76.48,145.33)
\path(76.48,145.33)(76.97,145.41)(77.44,145.49)(77.94,145.55)(78.44,145.63)(78.94,145.69)(79.44,145.75)(79.94,145.8)(80.44,145.86)(80.94,145.91)
\path(80.94,145.91)(81.47,145.94)(81.98,146.0)(82.0,146.0)
\drawpath{48.0}{140.0}{82.0}{140.0}
\path(50.0,68.0)(50.0,68.0)(50.15,67.68)(50.31,67.36)(50.47,67.04)(50.65,66.72)(50.84,66.41)(51.0,66.08)(51.18,65.76)(51.36,65.47)
\path(51.36,65.47)(51.54,65.16)(51.75,64.86)(51.95,64.55)(52.15,64.23)(52.34,63.93)(52.54,63.63)(52.75,63.33)(52.95,63.02)(53.18,62.72)
\path(53.18,62.72)(53.38,62.43)(53.61,62.13)(53.84,61.84)(54.06,61.54)(54.29,61.25)(54.52,60.95)(54.75,60.65)(55.0,60.36)(55.24,60.08)
\path(55.24,60.08)(55.47,59.79)(55.72,59.5)(55.97,59.22)(56.22,58.93)(56.49,58.65)(56.75,58.36)(57.02,58.09)(57.27,57.81)(57.54,57.52)
\path(57.54,57.52)(57.83,57.25)(58.11,56.97)(58.38,56.7)(58.65,56.43)(58.95,56.15)(59.24,55.88)(59.54,55.61)(59.83,55.34)(60.13,55.08)
\path(60.13,55.08)(60.43,54.81)(60.74,54.54)(61.04,54.27)(61.36,54.02)(61.68,53.75)(61.99,53.5)(62.31,53.24)(62.63,52.97)(62.97,52.72)
\path(62.97,52.72)(63.29,52.45)(63.63,52.2)(63.97,51.95)(64.3,51.7)(64.65,51.45)(65.0,51.2)(65.33,50.95)(65.69,50.7)(66.05,50.45)
\path(66.05,50.45)(66.43,50.22)(66.79,49.97)(67.15,49.72)(67.52,49.49)(67.9,49.25)(68.27,49.0)(68.65,48.77)(69.02,48.54)(69.4,48.29)
\path(69.4,48.29)(69.8,48.06)(70.19,47.83)(70.58,47.59)(70.99,47.36)(71.4,47.13)(71.79,46.9)(72.19,46.68)(72.61,46.45)(73.01,46.24)
\path(73.01,46.24)(73.44,46.0)(73.86,45.79)(74.3,45.56)(74.72,45.34)(75.15,45.13)(75.58,44.9)(76.01,44.7)(76.47,44.47)(76.91,44.27)
\path(76.91,44.27)(77.33,44.06)(77.8,43.84)(78.25,43.63)(78.69,43.4)(79.16,43.22)(79.62,43.0)(80.08,42.79)(80.55,42.59)(81.04,42.4)
\path(81.04,42.4)(81.51,42.2)(81.98,42.0)(82.0,42.0)
\drawpath{50.0}{52.0}{84.0}{52.0}
\path(54.0,140.0)(54.0,140.0)(54.11,140.13)(54.24,140.3)(54.36,140.46)(54.47,140.61)(54.61,140.75)(54.72,140.91)(54.84,141.05)(54.97,141.19)
\path(54.97,141.19)(55.11,141.33)(55.22,141.47)(55.36,141.61)(55.49,141.74)(55.61,141.86)(55.75,142.0)(55.88,142.11)(56.02,142.25)(56.15,142.36)
\path(56.15,142.36)(56.27,142.49)(56.4,142.6)(56.56,142.72)(56.68,142.83)(56.83,142.91)(56.97,143.02)(57.11,143.13)(57.25,143.25)(57.38,143.33)
\path(57.38,143.33)(57.52,143.44)(57.65,143.52)(57.81,143.63)(57.95,143.72)(58.09,143.8)(58.24,143.88)(58.38,143.97)(58.54,144.05)(58.68,144.13)
\path(58.68,144.13)(58.83,144.19)(58.97,144.27)(59.13,144.33)(59.27,144.41)(59.43,144.47)(59.59,144.52)(59.74,144.6)(59.88,144.66)(60.04,144.71)
\path(60.04,144.71)(60.2,144.77)(60.36,144.82)(60.52,144.86)(60.68,144.91)(60.84,144.94)(60.99,145.0)(61.15,145.02)(61.31,145.07)(61.47,145.1)
\path(61.47,145.1)(61.63,145.13)(61.79,145.16)(61.97,145.19)(62.13,145.22)(62.29,145.24)(62.47,145.25)(62.63,145.27)(62.79,145.27)(62.97,145.3)
\path(62.97,145.3)(63.13,145.3)(63.31,145.32)(63.47,145.33)(63.65,145.33)(63.83,145.33)(64.0,145.33)(64.18,145.32)(64.33,145.3)(64.52,145.3)
\path(64.52,145.3)(64.69,145.27)(64.88,145.27)(65.05,145.25)(65.24,145.25)(65.43,145.22)(65.61,145.19)(65.79,145.16)(65.97,145.13)(66.15,145.11)
\path(66.15,145.11)(66.33,145.08)(66.52,145.05)(66.69,145.0)(66.9,144.97)(67.08,144.91)(67.27,144.88)(67.44,144.83)(67.65,144.77)(67.83,144.72)
\path(67.83,144.72)(68.02,144.66)(68.22,144.61)(68.41,144.55)(68.61,144.5)(68.8,144.41)(69.0,144.36)(69.19,144.3)(69.4,144.22)(69.58,144.13)
\path(69.58,144.13)(69.79,144.07)(69.99,144.0)(70.0,144.0)
\path(54.0,134.0)(54.0,134.0)(54.22,134.0)(54.47,134.0)(54.7,134.0)(54.95,134.0)(55.18,134.0)(55.4,134.02)(55.65,134.02)(55.88,134.02)
\path(55.88,134.02)(56.11,134.02)(56.34,134.05)(56.56,134.07)(56.79,134.08)(57.0,134.1)(57.24,134.11)(57.45,134.13)(57.68,134.13)(57.9,134.16)
\path(57.9,134.16)(58.11,134.19)(58.34,134.21)(58.56,134.24)(58.77,134.25)(58.97,134.27)(59.2,134.3)(59.4,134.33)(59.61,134.36)(59.83,134.38)
\path(59.83,134.38)(60.04,134.41)(60.24,134.47)(60.45,134.5)(60.65,134.52)(60.86,134.57)(61.06,134.61)(61.25,134.63)(61.45,134.69)(61.65,134.72)
\path(61.65,134.72)(61.86,134.77)(62.04,134.82)(62.25,134.86)(62.43,134.91)(62.63,134.94)(62.83,135.0)(63.02,135.05)(63.2,135.1)(63.38,135.16)
\path(63.38,135.16)(63.58,135.21)(63.77,135.25)(63.95,135.32)(64.13,135.38)(64.3,135.44)(64.49,135.5)(64.66,135.55)(64.83,135.61)(65.02,135.66)
\path(65.02,135.66)(65.19,135.74)(65.38,135.8)(65.54,135.88)(65.72,135.94)(65.9,136.0)(66.05,136.08)(66.22,136.13)(66.4,136.22)(66.55,136.3)
\path(66.55,136.3)(66.72,136.38)(66.9,136.44)(67.05,136.52)(67.22,136.61)(67.38,136.69)(67.54,136.77)(67.69,136.85)(67.83,136.94)(68.0,137.02)
\path(68.0,137.02)(68.15,137.11)(68.3,137.19)(68.47,137.27)(68.61,137.36)(68.77,137.46)(68.91,137.55)(69.05,137.63)(69.19,137.74)(69.33,137.83)
\path(69.33,137.83)(69.5,137.91)(69.63,138.02)(69.77,138.13)(69.91,138.22)(70.05,138.33)(70.19,138.41)(70.33,138.52)(70.47,138.63)(70.58,138.75)
\path(70.58,138.75)(70.72,138.85)(70.86,138.96)(71.0,139.07)(71.13,139.16)(71.25,139.3)(71.38,139.41)(71.5,139.52)(71.63,139.63)(71.75,139.75)
\path(71.75,139.75)(71.86,139.88)(71.99,140.0)(72.0,140.0)
\path(58.0,58.0)(58.0,58.0)(58.18,58.06)(58.38,58.15)(58.59,58.22)(58.79,58.29)(58.99,58.36)(59.18,58.4)(59.38,58.49)(59.58,58.54)
\path(59.58,58.54)(59.77,58.59)(59.97,58.65)(60.15,58.7)(60.36,58.75)(60.56,58.79)(60.75,58.84)(60.95,58.88)(61.13,58.9)(61.34,58.95)
\path(61.34,58.95)(61.52,58.97)(61.72,59.0)(61.9,59.04)(62.11,59.06)(62.29,59.08)(62.49,59.09)(62.68,59.11)(62.86,59.11)(63.06,59.13)
\path(63.06,59.13)(63.25,59.13)(63.43,59.13)(63.63,59.13)(63.81,59.13)(64.0,59.13)(64.18,59.11)(64.38,59.11)(64.55,59.09)(64.75,59.08)
\path(64.75,59.08)(64.93,59.06)(65.11,59.04)(65.3,59.0)(65.49,58.99)(65.68,58.95)(65.86,58.9)(66.04,58.88)(66.22,58.84)(66.4,58.79)
\path(66.4,58.79)(66.58,58.75)(66.77,58.7)(66.94,58.65)(67.13,58.61)(67.3,58.54)(67.49,58.5)(67.66,58.43)(67.83,58.36)(68.02,58.29)
\path(68.02,58.29)(68.19,58.22)(68.38,58.15)(68.55,58.08)(68.75,58.0)(68.91,57.93)(69.08,57.84)(69.27,57.75)(69.44,57.65)(69.63,57.56)
\path(69.63,57.56)(69.79,57.47)(69.97,57.38)(70.15,57.27)(70.3,57.18)(70.5,57.06)(70.66,56.95)(70.83,56.84)(71.0,56.74)(71.18,56.61)
\path(71.18,56.61)(71.36,56.5)(71.52,56.36)(71.69,56.25)(71.86,56.11)(72.04,55.99)(72.19,55.84)(72.37,55.72)(72.55,55.58)(72.69,55.43)
\path(72.69,55.43)(72.87,55.29)(73.05,55.13)(73.22,54.99)(73.37,54.84)(73.55,54.68)(73.72,54.52)(73.87,54.36)(74.05,54.18)(74.19,54.02)
\path(74.19,54.02)(74.37,53.86)(74.54,53.68)(74.69,53.5)(74.87,53.33)(75.01,53.13)(75.19,52.95)(75.33,52.77)(75.51,52.58)(75.66,52.38)
\path(75.66,52.38)(75.83,52.18)(75.98,52.0)(76.0,52.0)
\path(56.0,52.0)(56.0,52.0)(56.08,51.84)(56.15,51.68)(56.24,51.52)(56.33,51.36)(56.4,51.22)(56.5,51.06)(56.59,50.9)(56.7,50.77)
\path(56.7,50.77)(56.79,50.63)(56.9,50.5)(57.0,50.36)(57.09,50.22)(57.2,50.08)(57.31,49.95)(57.4,49.81)(57.52,49.68)(57.63,49.56)
\path(57.63,49.56)(57.75,49.43)(57.88,49.31)(58.0,49.2)(58.11,49.08)(58.24,48.95)(58.36,48.84)(58.49,48.72)(58.61,48.61)(58.75,48.5)
\path(58.75,48.5)(58.88,48.4)(59.02,48.29)(59.15,48.2)(59.29,48.09)(59.43,48.0)(59.58,47.9)(59.72,47.79)(59.86,47.7)(60.02,47.61)
\path(60.02,47.61)(60.15,47.52)(60.31,47.43)(60.47,47.36)(60.63,47.27)(60.79,47.2)(60.95,47.11)(61.11,47.04)(61.27,46.95)(61.45,46.88)
\path(61.45,46.88)(61.61,46.81)(61.79,46.75)(61.95,46.68)(62.13,46.61)(62.31,46.56)(62.49,46.5)(62.68,46.43)(62.86,46.38)(63.04,46.31)
\path(63.04,46.31)(63.22,46.27)(63.4,46.22)(63.61,46.15)(63.79,46.11)(64.0,46.08)(64.19,46.04)(64.38,46.0)(64.58,45.95)(64.79,45.9)
\path(64.79,45.9)(65.0,45.88)(65.19,45.84)(65.41,45.81)(65.63,45.79)(65.83,45.75)(66.05,45.74)(66.27,45.72)(66.49,45.7)(66.72,45.68)
\path(66.72,45.68)(66.93,45.65)(67.15,45.63)(67.38,45.63)(67.61,45.61)(67.83,45.61)(68.08,45.59)(68.3,45.59)(68.55,45.59)(68.79,45.59)
\path(68.79,45.59)(69.04,45.59)(69.27,45.59)(69.52,45.59)(69.77,45.61)(70.02,45.61)(70.27,45.63)(70.52,45.63)(70.77,45.65)(71.04,45.68)
\path(71.04,45.68)(71.29,45.7)(71.55,45.72)(71.8,45.74)(72.08,45.75)(72.33,45.79)(72.62,45.81)(72.87,45.84)(73.16,45.88)(73.44,45.9)
\path(73.44,45.9)(73.72,45.95)(73.98,46.0)(74.0,46.0)
\path(54.0,140.0)(54.0,140.0)(54.18,140.03)(54.38,140.07)(54.59,140.11)(54.79,140.16)(54.99,140.19)(55.18,140.22)(55.38,140.27)(55.56,140.32)
\path(55.56,140.32)(55.75,140.35)(55.95,140.38)(56.15,140.44)(56.34,140.47)(56.52,140.5)(56.72,140.55)(56.9,140.6)(57.09,140.63)(57.27,140.66)
\path(57.27,140.66)(57.47,140.72)(57.65,140.75)(57.84,140.78)(58.02,140.83)(58.2,140.88)(58.38,140.91)(58.56,140.94)(58.75,141.0)(58.91,141.03)
\path(58.91,141.03)(59.09,141.07)(59.27,141.11)(59.45,141.16)(59.63,141.19)(59.81,141.22)(59.99,141.27)(60.15,141.32)(60.33,141.35)(60.5,141.38)
\path(60.5,141.38)(60.68,141.44)(60.84,141.47)(61.02,141.5)(61.18,141.55)(61.34,141.6)(61.52,141.63)(61.68,141.66)(61.86,141.72)(62.02,141.75)
\path(62.02,141.75)(62.18,141.78)(62.34,141.83)(62.5,141.88)(62.66,141.91)(62.83,141.94)(62.99,142.0)(63.15,142.03)(63.31,142.07)(63.47,142.11)
\path(63.47,142.11)(63.63,142.16)(63.77,142.19)(63.93,142.22)(64.09,142.27)(64.25,142.32)(64.4,142.35)(64.55,142.38)(64.7,142.44)(64.86,142.47)
\path(64.86,142.47)(65.0,142.5)(65.16,142.55)(65.3,142.6)(65.44,142.63)(65.59,142.66)(65.75,142.72)(65.88,142.75)(66.02,142.78)(66.18,142.83)
\path(66.18,142.83)(66.31,142.88)(66.45,142.91)(66.59,142.94)(66.74,143.0)(66.88,143.03)(67.02,143.07)(67.16,143.11)(67.3,143.16)(67.43,143.19)
\path(67.43,143.19)(67.56,143.22)(67.7,143.27)(67.83,143.32)(67.97,143.35)(68.09,143.38)(68.24,143.44)(68.36,143.47)(68.5,143.5)(68.63,143.55)
\path(68.63,143.55)(68.75,143.6)(68.88,143.63)(69.0,143.66)(69.13,143.72)(69.25,143.75)(69.38,143.78)(69.5,143.83)(69.63,143.88)(69.75,143.91)
\path(69.75,143.91)(69.86,143.94)(69.99,144.0)(70.0,144.0)
\path(54.0,134.0)(54.0,134.0)(54.15,134.11)(54.31,134.22)(54.47,134.35)(54.63,134.47)(54.79,134.58)(54.95,134.69)(55.11,134.8)(55.29,134.91)
\path(55.29,134.91)(55.45,135.02)(55.61,135.13)(55.77,135.24)(55.93,135.35)(56.11,135.44)(56.27,135.55)(56.43,135.66)(56.61,135.75)(56.77,135.86)
\path(56.77,135.86)(56.93,135.96)(57.11,136.05)(57.27,136.16)(57.43,136.25)(57.61,136.33)(57.77,136.44)(57.95,136.52)(58.11,136.61)(58.29,136.71)
\path(58.29,136.71)(58.45,136.8)(58.63,136.88)(58.79,136.97)(58.97,137.05)(59.15,137.13)(59.31,137.22)(59.49,137.3)(59.66,137.38)(59.84,137.46)
\path(59.84,137.46)(60.0,137.53)(60.18,137.61)(60.36,137.69)(60.54,137.75)(60.72,137.83)(60.88,137.91)(61.06,137.97)(61.24,138.05)(61.41,138.11)
\path(61.41,138.11)(61.59,138.17)(61.77,138.25)(61.95,138.3)(62.13,138.36)(62.31,138.42)(62.49,138.5)(62.68,138.55)(62.86,138.61)(63.04,138.66)
\path(63.04,138.66)(63.22,138.72)(63.4,138.77)(63.58,138.83)(63.75,138.88)(63.95,138.94)(64.13,138.99)(64.3,139.02)(64.5,139.08)(64.68,139.13)
\path(64.68,139.13)(64.86,139.16)(65.05,139.22)(65.24,139.25)(65.43,139.3)(65.61,139.33)(65.8,139.38)(65.99,139.41)(66.16,139.44)(66.36,139.49)
\path(66.36,139.49)(66.55,139.52)(66.74,139.55)(66.93,139.58)(67.11,139.61)(67.3,139.64)(67.5,139.67)(67.69,139.69)(67.88,139.72)(68.06,139.75)
\path(68.06,139.75)(68.27,139.77)(68.45,139.8)(68.65,139.82)(68.84,139.83)(69.04,139.86)(69.22,139.88)(69.43,139.88)(69.61,139.91)(69.81,139.91)
\path(69.81,139.91)(70.0,139.94)(70.2,139.94)(70.41,139.96)(70.59,139.97)(70.8,139.97)(71.0,139.97)(71.19,139.99)(71.4,139.99)(71.59,139.99)
\path(71.59,139.99)(71.8,139.99)(71.99,140.0)(72.0,140.0)
\path(58.0,58.0)(58.0,58.0)(58.11,57.88)(58.24,57.75)(58.36,57.63)(58.47,57.52)(58.61,57.4)(58.74,57.29)(58.86,57.18)(58.99,57.06)
\path(58.99,57.06)(59.11,56.95)(59.25,56.86)(59.38,56.75)(59.52,56.63)(59.65,56.54)(59.79,56.43)(59.93,56.33)(60.06,56.22)(60.2,56.13)
\path(60.2,56.13)(60.34,56.02)(60.49,55.93)(60.63,55.84)(60.77,55.74)(60.93,55.65)(61.06,55.54)(61.22,55.45)(61.36,55.36)(61.52,55.27)
\path(61.52,55.27)(61.66,55.18)(61.83,55.11)(61.97,55.02)(62.13,54.93)(62.29,54.84)(62.45,54.77)(62.61,54.68)(62.77,54.61)(62.93,54.52)
\path(62.93,54.52)(63.09,54.45)(63.25,54.38)(63.41,54.29)(63.59,54.22)(63.75,54.15)(63.91,54.08)(64.08,54.0)(64.25,53.93)(64.44,53.88)
\path(64.44,53.88)(64.61,53.81)(64.77,53.74)(64.95,53.68)(65.13,53.61)(65.31,53.56)(65.49,53.5)(65.68,53.43)(65.86,53.38)(66.04,53.31)
\path(66.04,53.31)(66.22,53.25)(66.41,53.2)(66.59,53.15)(66.77,53.09)(66.97,53.04)(67.16,53.0)(67.34,52.95)(67.55,52.9)(67.74,52.86)
\path(67.74,52.86)(67.94,52.81)(68.13,52.77)(68.33,52.72)(68.52,52.68)(68.72,52.65)(68.93,52.61)(69.13,52.56)(69.33,52.54)(69.54,52.5)
\path(69.54,52.5)(69.75,52.47)(69.94,52.43)(70.16,52.4)(70.36,52.36)(70.58,52.34)(70.79,52.31)(71.0,52.29)(71.22,52.25)(71.43,52.24)
\path(71.43,52.24)(71.65,52.2)(71.86,52.18)(72.08,52.16)(72.3,52.15)(72.52,52.13)(72.75,52.11)(72.98,52.09)(73.19,52.08)(73.43,52.06)
\path(73.43,52.06)(73.65,52.06)(73.87,52.04)(74.11,52.02)(74.33,52.02)(74.58,52.02)(74.8,52.0)(75.04,52.0)(75.27,52.0)(75.51,52.0)
\path(75.51,52.0)(75.76,52.0)(75.98,52.0)(76.0,52.0)(76.03,52.0)(76.14,52.0)(76.20,52.0)(76.25,52.0)
\path(56.0,52.0)(56.0,52.0)(56.18,51.95)(56.38,51.9)(56.59,51.86)(56.79,51.83)(56.99,51.79)(57.18,51.75)(57.38,51.7)(57.58,51.65)
\path(57.58,51.65)(57.77,51.61)(57.97,51.56)(58.16,51.52)(58.36,51.49)(58.56,51.43)(58.75,51.4)(58.95,51.34)(59.13,51.29)(59.34,51.25)
\path(59.34,51.25)(59.52,51.2)(59.72,51.15)(59.9,51.11)(60.11,51.06)(60.29,51.02)(60.49,50.97)(60.68,50.91)(60.86,50.86)(61.06,50.81)
\path(61.06,50.81)(61.25,50.77)(61.43,50.72)(61.63,50.66)(61.81,50.61)(62.0,50.56)(62.18,50.5)(62.38,50.45)(62.56,50.4)(62.75,50.34)
\path(62.75,50.34)(62.93,50.29)(63.11,50.24)(63.31,50.18)(63.49,50.13)(63.68,50.08)(63.86,50.02)(64.04,49.95)(64.22,49.9)(64.41,49.84)
\path(64.41,49.84)(64.58,49.79)(64.77,49.72)(64.94,49.66)(65.13,49.61)(65.3,49.54)(65.49,49.5)(65.66,49.43)(65.84,49.36)(66.02,49.31)
\path(66.02,49.31)(66.2,49.25)(66.38,49.18)(66.56,49.13)(66.75,49.06)(66.91,49.0)(67.09,48.93)(67.27,48.88)(67.44,48.81)(67.63,48.75)
\path(67.63,48.75)(67.8,48.68)(67.97,48.61)(68.15,48.54)(68.31,48.47)(68.5,48.41)(68.66,48.34)(68.83,48.27)(69.0,48.22)(69.19,48.15)
\path(69.19,48.15)(69.36,48.08)(69.52,48.0)(69.69,47.93)(69.86,47.86)(70.04,47.79)(70.2,47.72)(70.38,47.65)(70.55,47.59)(70.7,47.52)
\path(70.7,47.52)(70.88,47.43)(71.05,47.36)(71.22,47.29)(71.38,47.22)(71.55,47.15)(71.72,47.08)(71.88,47.0)(72.05,46.93)(72.2,46.84)
\path(72.2,46.84)(72.37,46.77)(72.54,46.7)(72.69,46.61)(72.87,46.54)(73.02,46.47)(73.19,46.38)(73.34,46.31)(73.51,46.22)(73.66,46.15)
\path(73.66,46.15)(73.83,46.06)(73.98,46.0)(74.0,46.0)
\end{picture}
\caption{Null curves and light-cones in the domains $0<t<\pi/2$ and
$\pi/2<t<\pi$.}
    \end{center}
\end{figure}

We next study the geometric behavior of the $t$-slices.

For any fixed $t\in \mathbb{R}$, it follows from (44) that the
induced metric of the $t$-slice reads
\begin{equation}\begin{array}{lll}
ds^2 & = & \eta_{22}dy^2+\eta_{33}dz^2\vspace{2mm}\\
& = & -e^{2x}\sin^2t[dy^2+(2+\sin y)^2dz^2].\end{array}
\end{equation}
When $t=k\pi\;(k\in \mathbb{N})$, the metric (48) becomes
$$ds^2=0.$$
This implies that the $t$-slice reduces to a point. On the other
hand, in the present situation, the metric (44) becomes
$$ds^2=2e^{x}(2+\sin y)dt^2.$$
When $t\neq k\pi\;(k\in \mathbb{N})$, (48) shows that the $t$-slice
is a three-dimensional cone-like manifold centered at $x=-\infty$.

{\em 3.2 Regular time-periodic space-times with non-vanishing
Riemann curvature tensor.} We next construct the regular
time-periodic space-times with non-vanishing Riemann curvature
tensor.

To do so, let
\begin{equation}
\left\{\begin{array}{l}
w(t)=\cos t,\vspace{2mm}\\
q(t)=\sin t,\vspace{2mm}\\
K(x,t)=e^x\sin t,\vspace{2mm}\\
N={\displaystyle -\frac{1}{(2+\sin y)^2}}.
\end{array}\right.
\end{equation}
Then, by (31),
$$V= \frac{\cos t\exp\left\{(2y-\cos y)\sin
t\right\}}{2+\sin y}.$$ Thus, in the present situation, we have the
following solution of the vacuum Einstein's field equations (2)
\begin{equation}
\widetilde{\eta}_{\mu\nu}= \left(
\begin{array}{cccc}
\widetilde{\eta}_{00} & \widetilde{\eta}_{01} & \widetilde{\eta}_{02} & 0 \\
\widetilde{\eta}_{01} & 0 & 0 & 0  \\
\widetilde{\eta}_{02} & 0 & \widetilde{\eta}_{22} & 0 \\
0 & 0 & 0 & \widetilde{\eta}_{33}\\
\end{array}\right),
\end{equation}
where
\begin{equation}
\left\{\begin{array}{lll}
\widetilde{\eta}_{00} & = & \displaystyle\frac{2e^x\cos^2t\exp\left\{(2y-\cos y)\sin t\right\}}{2+\sin y},\vspace{2mm}\\
\widetilde{\eta}_{01} & = & \displaystyle\frac{e^x\sin (2t)\exp\left\{(2y-\cos y)\sin t\right\}}{2(2+\sin y)},\vspace{2mm}\\
\widetilde{\eta}_{02} & = & e^x\left\{\sin t\cos
t-\displaystyle\frac{\cos t\cos y}{(2+\sin y)^2}\right\}\sin t\vspace{2mm}\\
& & \times\exp\left\{(2y-\cos y)\sin t\right\},\vspace{2mm}\\
\widetilde{\eta}_{22}& = & -e^{2x}\sin^2t,\vspace{2mm}\\
\widetilde{\eta}_{33}& = &
-\displaystyle\frac{e^{2x}\sin^2t}{(2+\sin y)^2}.
\end{array}\right.
\end{equation}
By (4),
\begin{equation}\begin{array}{lll}
\widetilde{\eta} & \stackrel{\triangle}{=} &
\det(\widetilde{\eta}_{\mu\nu})= -(\widetilde{\eta}_{01})^2\widetilde{\eta}_{22}\widetilde{\eta}_{33}\vspace{2mm}\\
& = & {\displaystyle -\frac{e^{6x+2(2y-\cos y)\sin
t}\sin^2(2t)\sin^4t}{4(2+\sin y)^4}.}\end{array}
\end{equation}

Introduce
$$\triangle (t,x,y)=6x+2(2y-\cos y)\sin t.$$
Thus, it follows from (52) that
\begin{equation}\widetilde{\eta}<0
\end{equation}
for $t\neq k\pi,\;k\pi+\pi/2\;\;(k\in \mathbb{N})$ and
$\triangle\neq -\infty$. It is obvious that the hypersurfaces $t=
k\pi,\;k\pi+\pi/2\;\;(k\in \mathbb{N})$ and $\triangle=\pm\infty$
are the singularities of the space-time described by (50) with (51).
As in Subsection 3.1, we can prove that  the hypersurfaces $t=
k\pi,\;k\pi+\pi/2\;\;(k\in \mathbb{N})$ are not essential (or say,
physical) singularities, these non-essential singularities
correspond to the event horizons of the space-time described by (50)
with (51).

Similar to Property 1, we have

\noindent{\bf Property 4} The solution (50) (in which
$(\widetilde{\eta}_{\mu\nu})$ is given by (51)) of the vacuum
Einstein's filed equations (2) is time-periodic.$\quad\blacksquare$

Similar to Property 2, we have

\noindent{\bf Property 5} The Lorentzian metric (50) (in which
$(\widetilde{\eta}_{\mu\nu})$ is given by (51)) describes a regular
space-time, this space-time has a non-vanishing Riemann curvature
tensor. $\quad\blacksquare$

\noindent{\bf Proof.} In the present situation, by (34)
\begin{equation} R_{\alpha\beta\mu\nu}=0,\quad\forall \;
\alpha\beta\mu\nu\neq 0202 \;\; {\rm or}\;\; 0303,
\end{equation}
while
\begin{equation}\begin{array}{lll}
R_{0202} & = & e^x (2+\sin y)\cos^2t\sin^2t \vspace{2mm}\\
&  & \times\exp\left\{(2y-\cos y)\sin t\right\},\end{array}
\end{equation}
and
\begin{equation}
R_{0303}=\frac{e^x\cos^2t\sin^2t\exp\left\{(2y-\cos y)\sin
t\right\}}{2+\sin y}.
\end{equation}
Property 5 follows from (54)-(56) directly. Thus the proof is
completed. $\quad\quad\quad\square$

In particular, when $t \neq k\pi,\;k\pi+\pi/2\;\;(k\in \mathbb{N})$,
it follows from (55) and (56) that
\begin{equation}
R_{0202},\; R_{0303}\longrightarrow \infty \quad {\rm as} \;\;
x+(2y-\cos y)\sin t\rightarrow \infty.
\end{equation}
However, a direct calculation gives
\begin{equation}
\mathbf{R}\triangleq
R^{\alpha\beta\gamma\delta}R_{\alpha\beta\gamma\delta}\equiv 0.
\end{equation}
Thus, we obtain

\noindent{\bf Property 6} The Lorentzian metric (50) (in which
$(\widetilde{\eta}_{\mu\nu})$ is given by (51)) does not contain any
essential singularity. These non-essential singularities consist of
the hypersurfaces $t= k\pi,\;k\pi+\pi/2\;\;(k\in \mathbb{N})$ and
$\triangle=\pm\infty$, in which the hypersurfaces $t=
k\pi,\;k\pi+\pi/2\;\;(k\in \mathbb{N})$ are the event horizons.
Moreover, the Riemann curvature tensor satisfies the properties (57)
and (58). $\quad\blacksquare$

We next analyze the singularity behavior of $\triangle=\pm\infty$.

\noindent{\bf Case 1:} Fixing $y\in \mathbb{R}$, we observe that
$$\triangle\rightarrow\pm\infty\Longleftrightarrow
x\rightarrow\pm\infty.$$ This situation is similar to the case
$x\rightarrow \pm\infty$ discussed in Subsection 3.1. That is to
say,  $x=+\infty$ corresponds to the event horizon, while, when
$x\rightarrow -\infty$, the space-time (50) with (51) degenerates to
a point.

\noindent{\bf Case 2:} Fixing $x\in \mathbb{R}$, we observe that
$$\triangle\rightarrow\pm\infty\Longleftrightarrow
y\rightarrow\pm\infty.$$ In the present situation, it holds that
$$t\neq k\pi\;(k\in \mathbb{N}).$$
Without loss of generality, we may assume that
$$\sin t>0. $$
For the case that $\sin t <0$, we have a similar discussion. Thus,
noting (57), we have
$$R_{0202},\; R_{0303}\longrightarrow \infty \quad {\rm as} \;\;
y\rightarrow \infty.$$ Moreover, by the definition of the event
horizon we can show that $y=+\infty$ is not a event horizon. On the
other hand, when $y\rightarrow -\infty$, the space-time (50) with
(51) degenerates to a point.

\noindent{\bf Case 3:} For the situation that
$x\rightarrow\pm\infty$ and $y\rightarrow\pm\infty$ simultaneously,
we have a similar discussion, here we omit the details.

For the space-time (50) with (51), the null curves and light-cones
are shown just as in Figure 1. On the other hand, for any fixed
$t\in \mathbb{R}$, the induced metric of the $t$-slice reads
\begin{equation}\begin{array}{lll}
ds^2 & = & \widetilde{\eta}_{22}dy^2+\widetilde{\eta}_{33}dz^2\vspace{2mm}\\
& = & -e^{2x}\sin^2t[dy^2+(2+\sin y)^{-2}dz^2].\end{array}
\end{equation}
Obviously, in the present situation, the $t$-slice possesses similar
properties shown in the last paragraph in Subsection 3.1.

In particular, if we take $(t,x,y,z)$ as the spherical coordinates
$(t,r,\theta,\varphi)$ with $t\in\mathbb{R},\;r\in
[0,\infty),\;\theta\in [0,2\pi),\;\varphi\in [-\pi/2,\pi/2]$, then
the metric (50) with (51) describes a regular time-periodic
space-time with non-vanishing Riemann curvature tensor. This
space-time does not contain any essential singularity, these
non-essential singularities consist of the hypersurfaces $t=
k\pi,\;k\pi+\pi/2\;\;(k\in \mathbb{N})$ which are the event
horizons. The Riemann curvature tensor satisfies (58) and
$$R_{0202},\; R_{0303}\longrightarrow \infty \quad {\rm as} \;\;
r \rightarrow \infty.$$ Moreover, when $t\neq k\pi\; (k\in
\mathbb{N})$, the $t$-slice is a three dimensional bugle-like
manifold with the base at $x=0$; while, when $t=k\pi\; (k\in
\mathbb{N})$, the $t$-slice reduces to a point.

{\em 3.3 Time-periodic space-times with physical singularities.}
This subsection is devoted to constructing the time-periodic
space-times with physical singularities.

To do so, let
\begin{equation} \label{eq:4.8}
\left\{\begin{array}{l}
w(t)=\cos t,\vspace{2mm}\\
q(t)=\sin t,\vspace{2mm}\\
K(x,t)=\displaystyle\frac{\sin t}{x^2},\vspace{2mm}\\
N=-\displaystyle\frac{1}{(2+\cos y)^2}.
\end{array}\right.
\end{equation}
Then, by (31) we have
$$V=\displaystyle\frac{\cos t\exp\left\{(2y+\sin y)\sin
t)\right\}}{2+\cos y}.$$ Thus, in the present situation, the
solution of the vacuum Einstein's field equations (2) in the
coordinates $(t,x,y,z)$ reads
\begin{equation}
ds^2=(dt,dx,dy,dz)(\hat{\eta}_{\mu\nu})(dt,dx,dy,dz)^T,
\end{equation}

where
\begin{equation}(\hat{\eta}_{\mu\nu})=\left( \begin{array}{cccc}
\hat{\eta}_{00} & \hat{\eta}_{01} & \hat{\eta}_{02} & 0\\ \hat{\eta}_{01} & 0 & 0 & 0 \\
\hat{\eta}_{02} & 0 & \hat{\eta}_{22} & 0\\
0 & 0 & 0 & \hat{\eta}_{33}\\
\end{array}\right),\end{equation}
in which
\begin{equation}
\left\{\begin{array}{lll}
\hat{\eta}_{00} & =& \displaystyle\frac{2\cos^2t\exp\left\{(\sin y+2y)\sin t\right\}}{(2+\cos y)x^2},\vspace{2mm}\\
\hat{\eta}_{01}& = & -\displaystyle\frac{\sin(2t)\exp\left\{(\sin y+2y)\sin t\right\}}{(2+\cos y)x^3},\vspace{2mm}\\
\hat{\eta}_{02} & = & \displaystyle\frac{\sin
t}{x^2}\left\{\frac{\cos t
\sin y}{(2+\cos y)^2}+\frac{\sin(2t)}{2}\right\}\times\vspace{2mm}\\
&  & \exp\left\{(\sin y+2y)\sin t\right\},\vspace{2mm}\\
\hat{\eta}_{22} & = & -\displaystyle\frac{\sin^2t}{x^4},\vspace{2mm}\\
\hat{\eta}_{33} & = & -\displaystyle\frac{\sin^2t}{(2+\cos y)^2x^4}.
\end{array}\right.
\end{equation}
By (4), we have
\begin{equation}\begin{array}{lll}\hat{\eta} & \stackrel{\triangle}{=} & \det(\hat{\eta}_{\mu\nu})=-
(\hat{\eta}_{01})^2\hat{\eta}_{22}\hat{\eta}_{33}\vspace{2mm}\\
& = & {\displaystyle -\frac{e^{2(2y+\sin y)\sin t}\sin^2(2t)\sin^4t
}{x^{14}(2+\cos y)^4}.}\end{array}
\end{equation}
It follows from (63) that
\begin{equation}\hat{\eta}<0
\end{equation}
for  $t\neq k\pi,\;k\pi+\pi/2\;\;(k\in \mathbb{N})$ and $x\neq 0$.
Obviously, the hypersurfaces $t= k\pi,\;k\pi+\pi/2\;\;(k\in
\mathbb{N})$ and $x=0$ are the singularities of the space-time
described by (61) with (62)-(63). As before, we can prove that the
hypersurfaces $t= k\pi,\;k\pi+\pi/2\;\;(k\in \mathbb{N})$ are not
essential (or, say, physical) singularities, and these non-essential
singularities correspond to the event horizons of the space-time
described by (61) with (62)-(63), however $x=0$ is an essential (or,
say, physical) singularity (see Property 8 below).

Similar to Property 1, we have

\noindent{\bf Property 7} The solution (61) (in which
$(\hat{\eta}_{\mu\nu})$ is given by (62) and (63)) of the vacuum
Einstein's field equations (2) is time-periodic. $\quad\blacksquare$

\noindent{\bf Proof.} In fact, the first equality in (63) implies
that
\begin{equation} \hat{\eta}_{00} >0 \quad {\rm for}\;\; t\neq
k\pi+\pi/2\quad (k\in \mathbb{N})\;\;{\rm and}\;\; x\neq 0.
\end{equation}
On the other hand, by direct calculations we have
\begin{equation}
\left|\begin{array}{cccc}
\hat{\eta}_{00} & \hat{\eta}_{01} \\
\hat{\eta}_{01}& 0
\end{array}\right|=-\hat{\eta}^2_{01}<0,
\end{equation}
\begin{equation}
\left|\begin{array}{cccc}
\hat{\eta}_{00} & \hat{\eta}_{01}&\hat{\eta}_{02}\\
\hat{\eta}_{01} & 0& 0 \\
\hat{\eta}_{02}& 0 &\hat{\eta}_{22}
\end{array}\right|=-\hat{\eta}_{01}^2\hat{\eta}_{22}
>0
\end{equation}
and
\begin{equation}\left|\begin{array}{cccc}
\hat{\eta}_{00}& \hat{\eta}_{01} &\hat{\eta}_{02} & 0  \\
\hat{\eta}_{01} & 0 & 0 & 0  \\
\hat{\eta}_{02} & 0& \hat{\eta}_{22} & 0 \\
0 & 0 & 0 & \hat{\eta}_{33}
\end{array}\right|=-\hat{\eta}_{01}^2\hat{\eta}_{22}\hat{\eta}_{33}<0
\end{equation}
for $t\neq k\pi,\;k\pi+\pi/2\;\;(k\in \mathbb{N})$ and $x\neq 0$.

The above discussion implies that the variable $t$ is a time
coordinate. Therefore, it follows from (63) that the Lorentzian
metric (61) is indeed a time-periodic solution of the vacuum
Einstein's field equations (2), where $(\hat{\eta}_{\mu\nu})$ is
given by (63). This proves Property 7. $\quad\quad\quad\square$

\noindent{\bf Property 8} When $t\neq k\pi,\;k\pi+\pi/2\;\;(k\in
\mathbb{N})$, for any fixed $y\in \mathbb{R}$ it holds that
\begin{equation}
R_{0202}\rightarrow +\infty\quad{\rm and}\quad R_{0303}\rightarrow
+\infty, \quad {\rm as} \;\; x\rightarrow 0.\end{equation} \hskip
8.3cm $\blacksquare$

\noindent{\bf Proof.} By direct calculations, we obtain from (35)
and (36) that
\begin{equation}
R_{0202}=\displaystyle\frac{(2+\cos y)\sin^2(2t)\exp\left\{(\sin
y+2y)\sin t\right\}}{4x^2},
\end{equation}
and
\begin{equation}
R_{0303}=\displaystyle\frac{\sin^2(2t)\exp\left\{(\sin(y)+2y)\sin
t\right\}}{4x^2(2+\cos y)}.
\end{equation}
(70) follows from (71) and (72) directly. The proof is finished.
$\quad\quad\quad\square$

On the other hand, a direct calculation yields
\begin{equation}
\mathbf{R}\triangleq
R^{\alpha\beta\gamma\delta}R_{\alpha\beta\gamma\delta}\equiv 0.
\end{equation}
Therefore, we have

\noindent{\bf Property 9}  The Lorentzian metric (61) describes a
time-periodic space-time, this space-time contains two kinds of
singularities: the hypersurfaces $t= k\pi,\;k\pi+\pi/2\;\;(k\in
\mathbb{N})$, which are non-essential singularities and correspond
to the event horizons, and $x=0$, which is an essential (or, say,
physical) singularity. $\quad\blacksquare$

We now analyze the behavior of the singularities of the space-time
characterized by (61) with (63).

By (64), we shall investigate the following cases: (a) $t=
k\pi,\;k\pi+\pi/2\;\;(k\in \mathbb{N})$; (b)
$y\rightarrow\pm\infty$;  (c) $x\rightarrow\pm\infty$; (d)
$x\rightarrow 0$.

\noindent {\bf Case a:} $t= k\pi,\;k\pi+\pi/2\;\;(k\in \mathbb{N})$.
According to the definition of the event horizon, the hypersurfaces
$t= k\pi,\;k\pi+\pi/2\;\;(k\in \mathbb{N})$ are the event horizons
of the space-time described by (61) with (63).

\noindent {\bf Case b:} $y\rightarrow\pm\infty$. Noting (64), in
this case we may assume that $t\neq k\pi\;(k\in \mathbb{N})$ (if
$t=k\pi$, then the situation becomes trivial). Without loss of
generality, we may assume that $\sin t>0$. Therefore, it follows
from (71) and (72) that, for any fixed $x\neq 0$ it holds that
\begin{equation}
R_{0202},\;\; R_{0303}\longrightarrow \infty \quad {\rm as} \;\; y
\rightarrow +\infty
\end{equation}
and
\begin{equation}
R_{0202},\;\; R_{0303}\longrightarrow 0 \quad {\rm as} \;\; y
\rightarrow -\infty.
\end{equation}
(74) implies that $y=+\infty$ is also a essential singularity, while
$y=-\infty$ is not because of (75).

\noindent {\bf Case c:} $x\rightarrow\pm\infty$. By (63), in this
case the space-time characterized by (61) reduces to a point.

\noindent {\bf Case d:} $x\rightarrow 0$. Property 8 shows that
$x=0$ is a physical singularity. This is the biggest difference
between the space-times presented in Subsections 3.1-3.2 and the one
given this subsection. In order to illustrate its physical meaning,
we take $(t,x,y,z)$ as the spherical coordinates
$(t,r,\theta,\varphi)$ with $t\in\mathbb{R},\;r\in
[0,\infty),\;\theta\in [0,2\pi),\;\varphi\in [-\pi/2,\pi/2]$. In the
coordinates $(t,r,\theta,\varphi)$, the metric (61) with (63)
describe a time-periodic space-time which possesses three kind of
singularities:

(i) \;\; $t\neq k\pi\;(k\in \mathbb{N})$: they are the event
horizons;

(ii) \; $r\rightarrow +\infty$: the space-time degenerates to a
point;

(iii) \,\,$r\rightarrow 0$: it is a physical singularity.

\noindent For the case (iii), in fact Property 8 shows that every
point in the set
$$\mathfrak{S}_B\stackrel{\triangle}{=}\{(t,r,\theta,\varphi)\,|\;r=0,
\;t\neq k\pi,\;k\pi+\pi/2\;(k\in \mathbb{N})\}$$ is a singular
point. Noting (34) and (70), we name the set of singular points
$\mathfrak{S}_B$ as a {\it quasi-black-hole}. Property 8 also shows
that the space-time (61) is not homogenous and not asymptotically
flat. This space-time perhaps has some new applications in cosmology
due to the recent WMAP data, since the recent WMAP data show that
our Universe exists anisotropy (see \cite{wmap}). This inhomogenous
property of the new space-time (61) may provide a way to give an
explanation of this phenomena.

We next investigate the physical behavior of the space-time (61).

Fixing $y$ and $z$, we get the induced metric
\begin{equation}
ds^2=\hat{\eta}_{00}dt^2+2\hat{\eta}_{01}dtdx.
\end{equation}
Consider the null curves in the $(t,x)$-plan defined by
\begin{equation}
\hat{\eta}_{00}dt^2+2\hat{\eta}_{01}dtdx=0.
\end{equation}
Noting (63) leads to
\begin{equation}
dt=0 \quad {\rm and} \quad \frac{dt}{dx}=-\frac{2\tan t}{x}.
\end{equation}
Let
\begin{equation}
\rho=2\ln{|x|}.
\end{equation}
Then the second equation in (78) becomes
\begin{equation}
\frac{dt}{d\rho}=-\tan t.
\end{equation}
Thus, in the ($t,\rho$)-plan the null curves and light-cones are
shown in Figure 1 in which $x$ should be replaced by $\rho$.

We now study the geometric behavior of the $t$-slices.

For any fixed $t\in \mathbb{R}$, the induced metric of the $t$-slice
reads
\begin{equation}
ds^2 =-\frac{\sin^2t}{x^4}[dy^2+(2+\cos y)^{-2}dz^2].
\end{equation}
When $t=k\pi\;(k\in \mathbb{N})$, the metric (81) becomes
$$ds^2=0.$$
This implies that the $t$-slice reduces to a point. On the other
hand, in this case the metric (61) becomes
$$ds^2=\frac{2}{(2+\cos y)x^2}dt^2.$$
When $t\neq k\pi\;(k\in \mathbb{N})$, (81) shows that the $t$-slice
is a three-dimensional manifold with cone-like singularities at
$x=\infty$ and $x=-\infty$, respectively. In particular, if we take
$(t,x,y,z)$ as the spherical coordinates $(t,r,\theta,\varphi)$,
then the induced metric (81) becomes
\begin{equation}
ds^2 =-\frac{\sin^2t}{r^4}[d\theta^2+(2+\cos\theta)^{-2}d\varphi^2].
\end{equation}
In this case the $t$-slice is a three-dimensional cone-like manifold
centered at $r=\infty$.

At the end of this subsection, we would like to emphasize that the
space-time (61) possesses a physical singularity, i.e., $x=0$ which
is named as a quasi-black-hole in this paper.

{\em 4. Summary and discussion.} In this paper we describe a new
method to find exact solutions of the Einstein's field equations
(1). Using our method, we can construct many interesting exact
solutions, in particular, the time-periodic solutions of the vacuum
Einstein's field equations. More precisely, we have constructed
three kinds of new time-periodic solutions of the vacuum Einstein's
field equations: the regular time-periodic solution with vanishing
Riemann curvature tensor, the regular time-periodic solution with
finite Riemann curvature tensor and the time-periodic solution with
physical singularities. We have also analyzed the singularities of
these new time-periodic solutions and investigate some new physical
phenomena enjoyed by these new space-times.

In particular, in the spherical coordinates $(t,r,\theta,\varphi)$
we construct a time-periodic space-time with essential
singularities. This space-time possesses an interesting and
important singularity which is named as a {\it quasi-black-hole}.
This space-time is inhomogenous and not asymptotically flat and can
perhaps be used to explain the phenomenon that our Universe exists
anisotropy from the recent WMAP data (see \cite{wmap}). We believe
some applications of these new space-times in modern cosmology and
general relativity can be expected.

The work of Kong was supported in part by the NSF of China (Grant
No. 10671124) and the Qiu-Shi Professor Fellowship from Zhejiang
University, China; the work of Liu was supported by the NSF and NSF
of China.

\end{document}